\documentclass[12pt]{article}
\usepackage{graphicx}
\usepackage{latexsym}
\usepackage{amscd}
\usepackage[cp1251]{inputenc}
\usepackage[english,russian]{babel}
\usepackage{amsmath,  eucal}
\usepackage{amssymb}
\usepackage{indentfirst}
\usepackage[small,centerlast]{caption2}
\usepackage{longtable}
\usepackage[dvipsnames]{color}
\usepackage{cite}
\graphicspath{{figure/}}


 {

\makeatletter
\renewcommand{\section}{\@startsection {section}{1}{\z@}%
                                   {-3.5ex \@plus -1ex \@minus -.2ex}%
                                   {2.3ex \@plus.2ex}%
                                   {\normalfont\Large\uppercase}}
\renewcommand{\subsection}{\@startsection{subsection}{2}{\z@}%
                                     {-3.25ex\@plus -1ex \@minus -.2ex}%
                                     {1.5ex \@plus .2ex}%
                                     {\normalfont\large\itshape}}
\renewcommand{\subsubsection}{\@startsection{subsubsection}{3}{1em}%
                                     {-3.25ex\@plus -1ex \@minus -.2ex}%
                                     {-1.5em \@plus .2em}%
                                     {\normalfont\normalsize\bfseries}}

\makeatother

\begin{document}
\renewcommand{\refname}{\begin{center}\bf REFERENCES\end{center}}
\newcommand{\mc}[1]{\mathcal{#1}}
\newcommand{\E}{\mc{E}}
\thispagestyle{empty} \large
\renewcommand{\abstractname}{\,}

 \begin{center}
{\textbf{Nonlinear longitudinal current in degenerate plasma,
arising under the influence of the transversal electromagnetic field}}
\medskip
\end{center}

\begin{center}
  \bf A. V. Latyshev\footnote{$avlatyshev@mail.ru$} and
  A. A. Yushkanov\footnote{$yushkanov@inbox.ru$}
\end{center}\medskip

\begin{abstract}
Kinetic Vlasov-Boltzmann  equation for degenerate collisional plasmas
with integral of collisions of relaxation type BGK (Bhatnagar,
Gross and Krook) is used. Square-law expansion  on
size of intensity of electric field for kinetic
equation, Lorentz's force and integral of collisions is considered.

It is shown, that nonlinearity leads to generation of the longitudinal
electric current directed along a wave vector.
Longitudinal current is perpendicular to the known transversal
classical current received at the linear analysis.

The case of small values of wave number is considered.

When frequency of collisions tends to the zero, all received
results for collisional pass plasmas in
corresponding results for collisionless plasmas.

Graphic research of the real and imaginary part
current density is carried out.
\end{abstract}

\section*{\bf Введение}

В настоящей работе выводятся формулы для вычисления
электрического тока в классической вырожденной плазме.

При решении уравнения Власова, описывающего поведение
вырожденной столкновительной плазмы, мы учитываем
величины, пропорциональные квадрату напряженности внешнего электрического
поля. Эти величины мы учитываем в разложении
функции распределения, в разложении самосогласованного
электромагнитного поля, а
также в разложении интеграла столкновений .

При таком нелинейном подходе оказалось, что электрический ток
имеет две ненулевые компоненты. Одна компонента электрического
тока направлена вдоль напряженности электрического поля. Эта
компонента в точности та же самая, что и при линейном анализе.
Это поперечный ток. Следовательно, в линейном анализе мы
получаем хорошо известное выражение поперечного электрического
тока.

Вторая ненулевая компонента электрического тока имеет второй
порядок малости по отношению к напряженности электрического
поля. Вторая компонента электрического тока направлена вдоль
волнового вектора. Этот ток ортогонален первой компоненте. Это
продольный ток.

К появлению продольного тока приводит нелинейный анализ
взаимодействия электромагнитного поля с плазмой.

Нелинейные эффекты в плазме исследуются уже длительное время
\cite{Gins} -- \cite{Lat1}.

В работе \cite{Zyt} был исследован нелинейный ток, в частности,
в вопросах вероятности распадных процессов.
Заметим, что в работе \cite{Zyt2} указывается существование
нелинейного тока вдоль волнового вектора (см. формула (2.9) из \cite{Zyt2}).

Квантовая плазма изучалась в работах \cite{Lat1} - \cite{Lat8}.
Диэлектрическая проницаемость плазмы в квантовой
столкновительной плазме начала изучаться в работе
\cite{Mermin}. Затем исследование квантовой столкновительной
плазмы продолжалось в наших работах \cite{Lat2} - \cite{Lat5}.
В этих работах изучалась диэлектрическая проницаемость квантовой
столновительной плазмы с произвольной переменной частотой
столкновений электронов с частицами плазмы.

В работах \cite{Lat7} - \cite{Lat9}
было исследовано генерирование продольного тока поперечным
электромагнитным полем в классической и квантовой плазме
Ферми---Дирака \cite{Lat7}, в максвелловской плазме \cite{Lat8}
и в вырожденной плазме \cite{Lat9}.
В работе \cite{Lat10} изучалось генерирование продольного тока
поперечным электромагнитным полем в классической
столкновительной плазме Ферми---Дирака при произвольной
температуре (т.е. при произвольной степени вырождения электронного
газа).

В настоящей работе выведены формулы для вычисления
электрического тока в классической вырожренной плазме.

\section{\bf Решение уравнения Власова}

Возьмем уравнение Власова, описывающее поведение
столкновительной плазмы с интегралом столкновений БГК (Бхатнагар, Гросс
и Крук)\medskip
$$
\dfrac{\partial f}{\partial t}+\mathbf{v}\dfrac{\partial f}{\partial
\mathbf{r}}+
e\bigg(\mathbf{E}+
\dfrac{1}{c}[\mathbf{v},\mathbf{H}]\bigg)
\dfrac{\partial f}{\partial\mathbf{p}}=\nu(f^{(0)}-f).
\eqno{(1.1)}
$$\medskip

В уравнении (1.1) $f$ -- функция распределения электронов плазмы,
${\bf E}, {\bf H}$ -- компоненты электромагнитного поля,
$c$ -- скорость света,
$\nu$ -- эффективная частота столкновений электронов с частицами
плазмы,
$f^{(0)}=f_{eq}({\bf r},v)$ (eq $\equiv$ equilib\-rium)
-- локально равновесное распределение Ферми,\medskip
$$
f_{eq}({\bf r},v,t)=\Theta(\E_0({\bf r},t)-\E),
$$
\medskip
где $\Theta(x)$ -- единичная ступенька Хэвисайда,
$$
\Theta(x)=1,\; x>0;\qquad \Theta(x)=0,\; x<0,
$$

$$
\E_0({\bf r},t)=\dfrac{{mv_0^2({\bf r},t)}}{2}
$$
-- возмущенная энергия электронов на поверхности Ферми,
$v_0({\bf r},t)$ --  возмущенная скорость электронов на поверхности
Ферми, и  $\E=mv^2/2$ --  энергия электронов,
${\bf p}_0({\bf r},t)=m{\bf v}_0({\bf r},t)$ -- возмущенный импульс
электронов на поверхности Ферми.

Обозначим
$$
{\bf P}=\dfrac{{\bf p}}{p_0}=\dfrac{{\bf v}}{v_0},
$$
${\bf P}$  -- безразмерный импульс электронов,  $p_0=mv_0$
импульс электронов на поверхности Ферми,
$v_0$ -- скорость электронов на поверхности Ферми.

Будем считать, что в плазме существует электромагнитное поле,
представляющее собой бегущую гармоническую волну
$$
{\bf E}={\bf E}_0e^{i({\bf kr}-\omega t)}, \qquad
{\bf H}={\bf H}_0e^{i({\bf kr}-\omega t)}.
$$

Электрическое и магнитное поля связаны с векторным потенциалом
равенствами
$$
\mathbf{E}=-\dfrac{1}{c}\dfrac{\partial \mathbf{A}}{\partial t},
\;\qquad
\mathbf{H}={\rm rot} \mathbf{A}.
$$

Будем предполагать, что волновой вектор ортогонален векторному
потенциалу электромагнитного поля:
$$
{\bf k}{\bf A}({\bf r},t)=0.
$$

Для определенности будем считать, что волновой вектор направлен
вдоль оси $x $, а электрическое поле направлено вдоль оси $y $,
т.e.
$$
{\bf k}=k(1,0,0), \qquad {\bf E}=E_y(x,t)(0,1,0).
$$

Следовательно,
$$
\mathbf{E}=-\dfrac{1}{c}\dfrac{\partial \mathbf{A}}{\partial t}
=\dfrac{i\omega}{c}\mathbf{A},
$$
$$
{\bf H}=\dfrac{ck}{\omega}E_y\cdot(0,0,1),\qquad
{\bf [v,H}]=\dfrac{ck}{\omega}E_y\cdot (v_y,-v_x,0),
$$
$$
e\bigg(\mathbf{E}+\dfrac{1}{c}[\mathbf{v},\mathbf{H}]\bigg)
\dfrac{\partial f}{\partial\mathbf{p}}=
\dfrac{e}{\omega}E_y\Big[kv_y\dfrac{\partial f}{\partial p_x}+
(\omega-kv_x)\dfrac{\partial f}{\partial p_y}\Big],
$$
а также
$$
[\mathbf{v,H}]\dfrac{\partial f_0}{\partial \mathbf{p}}=0,\quad
\text{because}\quad
\dfrac{\partial f_0}{\partial \mathbf{p}}\sim \mathbf{v}.
$$

Преобразуем локально равновесную функцию распределения
электронов в вырожденной плазме. Имеем:
$$
f_{eq}(x,v,t)=\Theta\Big[\dfrac{v_0^2(x,t)-v^2}{v_0^2}\Big]=
\Theta\Big[P_0^2(x,t)-P^2\Big]=
$$
$$
=\Theta\Big[P_0(x,t)-P\Big]=f_{eq}(P,x,t).
$$
Здесь
$$
{\bf P}_0(x,t)=\dfrac{{\bf p_0}(x,t)}{p_0}=\dfrac{{\bf v_0}(x,t)}{v_0}
$$
-- безразмерный возмущенный импульс (скорость) электронов.

Рассмотрим линеаризацию функции $f_{eq}(P,x,t)$ относительно
поверхности Ферми, полагая
$$
P_0(x,t)=P_0+\delta P_0(x,t).
$$

Здесь $P_0=1$, ибо на поверхности Ферми $P_0=p_0/p_0=1$.

Проведем линеаризацию локально равновесной функции распределения
$$
f_{eq}(P,x)=\Theta\Big[P_0(x,t)-P\Big]=\Theta\Big[1-P+\delta P_0(x,t)\Big]=
$$
$$
=f_0(P)+\delta(1-P)\delta P_0(x,t),
$$
где $\delta(x)$ -- дельта--функция Дирака,
$$
f_0(P)=\Theta(1-P).
$$

Уравнение  (1.1) теперь может быть переписано в следующем виде
$$
\dfrac{\partial f}{\partial t}+v_x\dfrac{\partial f}{\partial x}+
\dfrac{eE_y}{\omega}\Big[kv_y\dfrac{\partial f}{\partial p_x}+
(\omega-kv_x)\dfrac{\partial f}{\partial p_y}\Big]+\nu f=
$$
$$
=\nu f_0(P)+\nu \delta(1-P) \delta P_0(x,t).
\eqno{(1.2)}
$$

Будем искать решение уравнения (1.2) в виде
$$
f=f_0(P)+f_1+f_2,
\eqno{(1.3)}
$$
где
$$
f_1\sim E_y\sim e^{i(kx-\omega t)},
$$
$$
f_2\sim E_y^2\sim e^{2i(kx-\omega t)}.
$$

Величину $ \delta P(x,t) $ найдем из закона сохранения числа частиц
$$
\int (f_{eq}-f)\dfrac{2d^3p}{(2\pi \hbar)^2}=0.
$$

Из этого закона сохранения находим
$$
\delta P_0(x,t)\int \delta(1-P)\dfrac{2d^3p}{(2\pi \hbar)^2}=\int
[f-f_0(P)]\dfrac{2d^3p}{(2\pi \hbar)^2}.
$$
Из этого уравнения находим, что
$$
\delta P_0(x,t)=\dfrac{\displaystyle\int [f-f_0(P)] d^3P}
{\displaystyle\int \delta(1-P)d^3P}.
$$

Заметим, что
$$
\int \delta (1-P)d^3P=\int\limits_{-1}^{1}d\mu
\int\limits_{0}^{2\pi}d\chi
 \int\limits_{0}^{\infty}\delta (1-P)P^2dP=$$$$=
4\pi \int\limits_{0}^{\infty}\delta (1-P)P^2dP=4\pi.
$$

Следовательно
$$
\delta P_0(x,t)=\dfrac{1}{4\pi}\int [f-f_0(P)] d^3P.
$$

Теперь уравнение (1.2) может быть преобразовано к интегральному
уравнению
$$
\dfrac{\partial f}{\partial t}+v_x\dfrac{\partial f}{\partial x}+
\nu f=\nu f_0(P)-\dfrac{eE_y}{\omega}\Big[kv_y\dfrac{\partial f}{\partial p_x}+
(\omega-kv_x)\dfrac{\partial f}{\partial p_y}\Big]+
$$
$$
+\nu \delta(1-P)\dfrac{1}{4\pi}\int [f-f_0(P)] d^3P.
\eqno{(1.4)}
$$

Будем действовать методом последовательных приближений, считая
малым параметром величину напряженности электрического поля.
Тогда уравнение (1.4) с помощью (1.3) эквивалентно следующим
уравнениям
$$
\dfrac{\partial f_1}{\partial t}+
v_x\dfrac{\partial f_1}{\partial x}+\nu f_1=
$$
$$
=-\dfrac{eE_y}{\omega}\Bigg[kv_y\dfrac{\partial f_0}{\partial p_x}+
(\omega-kv_x)\dfrac{\partial f_0}{\partial p_y}\Bigg]
+\nu \delta(1-P)\dfrac{1}{4\pi}\int f_1 d^3P.
\eqno{(1.5)}
$$ \bigskip
и
$$
\dfrac{\partial f_2}{\partial t}+
v_x\dfrac{\partial f_2}{\partial x}+\nu f_2=
$$
$$
=-\dfrac{eE_y}{\omega}\Bigg[kv_y\dfrac{\partial f_1}{\partial p_x}+
(\omega-kv_x)\dfrac{\partial f_1}{\partial p_y}\Bigg]
+\nu \delta(1-P)\dfrac{1}{4\pi}\int f_2 d^3P.
\eqno{(1.6)}
$$ \bigskip

Из уравнения (1.5) получаем, что
$$
(\nu-i\omega+ikv_x)f_1=
$$
$$
=-\dfrac{eE_y}{\omega}
\Bigg[kv_y\dfrac{\partial f_0}{\partial p_x}+
(\omega-kv_x)\dfrac{\partial f_0}{\partial p_y}\Bigg]
+\nu \delta(1-P)A_1.
$$

Здесь
$$
A_1=\dfrac{1}{4\pi}\int f_1 d^3P.
\eqno{(1.7)}
$$

Введем безразмерные параметры
$$
\Omega=\dfrac{\omega}{k_0v_0},\qquad y=\dfrac{\nu}{k_0v_0},
\qquad q=\dfrac{k}{k_0}.
$$

Здесь $q $ -- безразмерное волновое число,
$k_0 =\dfrac {mv_0} {\hbar} $ волновое число Ферми, $ \Omega $
-- безразмерная частота столкновений колебаний электромагнитного поля.

В предыдущем уравнении перейдем к безразмерным параметрам
$$
i(qP_x-z)f_1=$$$$=-\dfrac{eE_y}{\Omega k_0p_0v_0}
\Bigg[qP_y\dfrac{\partial f_0}{\partial P_x}+
(\Omega-qP_x)\dfrac{\partial f_0}{\partial P_y}\Bigg]
+y \delta(1-P)A_1.
\eqno{(1.8)}
$$

Здесь
$$
z=\Omega+iy=\dfrac{\omega+iy}{k_0v_0}.
$$

Заметим, что
$$
\dfrac{\partial f_0}{\partial P_x}\sim P_x,\qquad
\dfrac{\partial f_0}{\partial P_y}\sim P_y.
$$

Следовательно
$$
\Bigg[qP_y\dfrac{\partial f_0}{\partial P_x}+
(\Omega-qP_x)\dfrac{\partial f_0}{\partial P_y}\Bigg]=
\Omega\dfrac{\partial f_0}{\partial P_y}.
$$

Теперь из уравнения (1.8) найдем, что
$$
f_1=\dfrac{ieE_y}{k_0p_0v_0}\cdot\dfrac{\partial f_0/\partial P_y}
{qP_x-z}-iy\cdot\dfrac{\delta(1-P)}{qP_x-z}A_1.
\eqno{(1.9)}
$$

Подставим  (1.9) в уравнение (1.7). Получим равенство
$$
A_1\Bigg(1+iy\int \dfrac{\delta(1-P)d^3P}
{qP_x-z}\Bigg)=\dfrac{ieE_y}{k_0p_0v_0}\int
\dfrac{\partial f_0/\partial P_y}{qP_x-z}d^3P.
$$

Легко видеть, что интеграл в правой части этого равенства равен
нулю. Следовательно, $A_1=0$.

Значит, согласно
(1.9) функция $f_1$ построена и определяется равенством
$$
f_1=\dfrac{ieE_y}{k_0p_0v_0}\cdot\dfrac{\partial f_0/\partial P_y}
{qP_x-z}.
\eqno{(1.10)}
$$

Во втором приближении мы подставим $f_1 $  согласно (1.10)
в уравнение (1.6).

Получаем уравнение
$$
(\nu-2i\omega+2ikv_x)f_2=$$$$
-\dfrac{ie^2E_y^2}{k_0p_0v_0\omega}\Big[kv_y\dfrac{\partial}{\partial p_x}
\Big(\dfrac{\partial f_0/\partial P_y}{qP_x-z}\Big)+
(\omega-kv_x)\dfrac{\partial }
{\partial p_y}\Big(\dfrac{\partial f_0/\partial
P_y}{qP_x-z}\Big)\Big]+
$$
$$
+\nu \delta(1-P)A_2.
$$

Здесь
$$
A_2=\dfrac{1}{4\pi}\int f_2d^3P.
\eqno{(1.11)}
$$

Перейдем в этом уравнении к безразмерным параметрам. Получаем
уравнение
$$
2i(qP_x-x-\dfrac{iy}{2})f_2=$$$$=-\dfrac{ie^2E_y^2}{\Omega k_0^2p_0^2v_0^2}
\Big[qP_x\dfrac{\partial}{\partial P_x}
\Big(\dfrac{\partial f_0/\partial P_y}{qP_x-z}\Big)+
(\Omega-qP_x)\dfrac{\partial }
{\partial P_y}\Big(\dfrac{\partial f_0/\partial
P_y}{qP_x-z}\Big)\Big]+
$$
$$
+y\delta(1-P)A_2.
$$

Обозначим
$$
z'=\Omega+\dfrac{iy}{2}=\dfrac{\omega}{k_0v_0}+i\dfrac{\nu}{2k_0v_0}=
\dfrac{\omega+i \nu/2}{k_0v_0}.
$$
Из последнего уравнения находим
$$
f_2=-\dfrac{e^2E_y^2}{2k_0^2p_0^2v_0^2 \Omega}
\Bigg[qP_y\dfrac{\partial}{\partial P_x}
\Big(\dfrac{\partial f_0/\partial P_y}{qP_x-z}\Big)+
\dfrac{\Omega-qP_x}{qP_x-z}\dfrac{\partial^2f_0}{\partial P_y^2}\Bigg]
\dfrac{1}{qP_x-z'}-
$$
$$
-\dfrac{iy}{2}\cdot\dfrac{\delta(1-P)}{qP_x-z'}A_2.
\eqno{(1.12)}
$$

Для нахождения $A_2$ подставим (1.12) в (1.11). Из полученного уравнения
находим $A_2$
$$
A_2=-\dfrac{e^2E_y^2}{2k_0^2p_0^2v_0^2\Omega}\cdot\dfrac{J_1}
{4\pi+\dfrac{iy}{2}J_0}.
$$

Здесь
$$
J_0=\int\dfrac{\delta(1-P)d^3P}{qP_x-z'}=
2\pi \int\limits_{-1}^{1}\dfrac{d\tau}{q\tau-z'}=
\dfrac{2\pi}{q}\ln\dfrac{z'-q}{z'+q},
$$
$$
J_1=\int\Bigg[qP_y\dfrac{\partial}{\partial P_x}
\Big(\dfrac{\partial f_0/\partial P_y}{qP_x-z}\Big)+
\dfrac{\Omega-qP_x}{qP_x-z}\dfrac{\partial^2f_0}{\partial P_y^2}\Bigg]
\dfrac{d^3P}{qP_x-z'}.
$$

Подставляя  $A_2$ в (1.12), окончательно находим функцию $f_2$
в явной форме
$$
f_2=-\dfrac{e^2E_y^2}{2k_0^2p_0^2v_0^2 \Omega}
\Bigg[qP_y\dfrac{\partial}{\partial P_x}
\Big(\dfrac{\partial f_0/\partial P_y}{qP_x-z}\Big)+
\dfrac{\Omega-qP_x}{qP_x-z}\dfrac{\partial^2f_0}{\partial P_y^2}\Bigg]
\dfrac{1}{qP_x-z'}+
$$
$$
+\gamma\dfrac{e^2E_y^2}{2k_0^2p_0^2v_0^2 \Omega}
\cdot\dfrac{\delta(1-P)}{qP_x-z'},
\eqno{(1.13)}
$$
где
$$
\gamma=\dfrac{(iy/2)J_1}{4\pi+(iy/2)J_0}.
\eqno{(1.14)}
$$

\section*{\bf 2 \quad Плотность электрического тока}

Найдем плотность электрического тока
$$
\mathbf{j}=e\int \mathbf{v}f \dfrac{2d^3p}{(2\pi\hbar)^3}.
\eqno{(2.1)}
$$

Из равенств (1.4) -- (1.6) видно, что вектор плотности тока имеет две
ненулевые компоненты
$$
\mathbf{j}=(j_x,j_y,0).
$$

Здесь $j_y$ -- плотность поперечного тока
$$
j_y=e\int v_yf \dfrac{2d^3p}{(2\pi\hbar)^3}=
e\int v_yf_1 \dfrac{2d^3p}{(2\pi\hbar)^3}.
$$

Этот ток направлен вдоль электрического поля, его плотность
определяется только первым приближением функции распределения.

Второе приближение функции распределения не вносит вклад в
плотность этого тока.

Плотность поперечного тока определяется равенством
$$
j_y=\dfrac{2ie^2p_0^2}{(2\pi\hbar)^3k_0}E_y(x,t)
\int\dfrac{(\partial f_0/\partial P_y)P_y}{qP_x-z}d^3P.
$$

Этот ток пропорционален первой степени величины напряженности
электрического поля.

Для плотности продольного тока в соответствии с его определением
имеем
$$
j_x=e\int v_xf\dfrac{2d^3p}{(2\pi\hbar)^3}=
e\int v_xf_2\dfrac{2d^3p}{(2\pi\hbar)^3}=
\dfrac{2ev_0p_0^3}{(2\pi\hbar)^3}\int P_xf_2d^3P.
$$

С помощью (1.6) отсюда получаем, что
$$
j_x=\dfrac{e^3E_y^2m}{(2\pi\hbar)^3k_0^2\Omega}\Bigg[-\int
\Bigg[qP_y\dfrac{\partial}{\partial P_x}
\Big(\dfrac{\partial f_0/\partial P_y}{qP_x-z}\Big)+
\dfrac{x-qP_x}{qP_x-z}\cdot\dfrac{\partial^2f_0}{\partial P_y^2}\Bigg]
\dfrac{P_xd^3P}{qP_x-z'}+
$$
$$
+\gamma\int\dfrac{P_x \delta(1-P)d^3P}{qP_x-z'}
\Bigg].
\eqno{(2.2)}
$$

В интеграле от второго слагаемого из квадратной скобки
внутренний интеграл по $P_y $ равен нулю:
$$
\int\limits_{-\infty}^{\infty}\dfrac{\partial^2f_0}{\partial P_y^2}dP_y
=\dfrac{\partial f_0}{\partial P_y}\Bigg|_{P_y=-\infty}^{P_y=+\infty}=0.
$$

В первом интеграле из квадратной скобки (2.2) внутренний интеграл $P_x $
вычисляется по частям
$$
\int\limits_{-\infty}^{\infty}\dfrac{\partial}{\partial P_x}
\Big(\dfrac{\partial f_0/\partial P_y}{qP_x-z}\Big)
\dfrac{P_xdP_x}{qP_x-z'}=z' \int\limits_{-\infty}^{\infty}
\dfrac{\partial f_0/\partial P_y}{(qP_x-z)(qP_x-z')^2}dP_x.
$$

Следовательно, равенство (2.2) упрощается
$$
j_x=\dfrac{e^3E_y^2m}{(2\pi\hbar)^3k_0^2\Omega}\Bigg[-z'q\int
\dfrac{P_y(\partial f_0/\partial P_y)d^3P}{(qP_x-z)(qP_x-z')^2}+
$$
$$
+\gamma\int\dfrac{P_x \delta(1-P)d^3P}{qP_x-z'}\Bigg].
$$

Внутренний интеграл по переменной $P_y $ мы вычислим по частям
$$
\int\limits_{-\infty}^{\infty}P_y\dfrac{\partial f_0}{\partial P_y}dP_y=
P_yf_0\Bigg|_{P_y=-\infty}^{P_y=+\infty}-
\int\limits_{-\infty}^{\infty}f_0(P)dP_y=
-\int\limits_{-\infty}^{\infty}f_0(P)dP_y.
$$

Следовательно, выражением для продольного тока приводит к
выражению
$$
j_x=\dfrac{e^3E_y^2m}{(2\pi\hbar)^3k_0^2\Omega}\Bigg[z'q\int
\dfrac{f_0(P)d^3P}{(qP_x-z)(qP_x-z')^2}+
$$
$$
+\gamma\int\dfrac{P_x \delta(1-P)}{qP_x-z'}d^3P\Bigg].
\eqno{(2.3)}
$$

Внутренний интеграл в плоскости $ (P_y, P_z) $ вычислим в полярных
координатах
$$
\int\dfrac{f_0(P)d^3P}{(qP_x-z')^2(qP_x-z)}=
$$
$$
=\int\limits_{-1}^{1}
\dfrac{dP_x}{(qP_x-z')^2(qP_x-z)}
\iint\limits_{P_y^2+P_z^2<1-P_x^2}dP_ydP_z=
$$
$$
=\pi\int\limits_{-1}^{1}
\dfrac{(1-P_x^2)dP_x}{(qP_x-z)(qP_x-z')^2},
$$
так как
$$
\iint\limits_{P_y^2+P_z^2<1-P_x^2}dP_ydP_z=\pi(1-P_x^2).
$$

Кроме того
$$
\int\dfrac{P_x \delta(1-P^2)d^3P}{qP_x-z'}=
\int\limits_{-1}^{1}\int\limits_{0}^{2\pi}\int\limits_{0}^{\infty}
\dfrac{\delta(P-1)\mu P^3 d\mu d\chi dP}{q \mu P-z'}=
$$
$$
=2\pi\int\limits_{-1}^{1}\dfrac{\mu d\mu}{q \mu-z'}=
2\pi\Bigg[\dfrac{2}{q}+\dfrac{z'}{q^2}\ln\dfrac{z'-q}{z'+q}\Bigg].
$$

Равенство (2.3) сводится к одномерному интегралу
$$
j_x=\dfrac{e^3E_y^2m}{(2\pi\hbar)^3k_0^2\Omega}\Bigg[z'q\pi\int\limits_{-1}^{1}
\dfrac{(1-P_x^2)dP_x}{(qP_x-z)(qP_x-z')^2}+
\gamma 2 \pi\int\limits_{-1}^{1}\dfrac{\tau d\tau}{q\tau-z'}\Bigg],
$$
или
$$
j_x=\dfrac{\pi e^3E_y^2mq}{(2\pi\hbar)^3k_0^2\Omega}\Big[z'J_{12}
+\gamma\dfrac{2}{q}J_{02}\Big],
\eqno{(2.4)}
$$
где
$$
J_{12}=\int\limits_{-1}^{1}
\dfrac{(1-\tau^2)d\tau}{(q\tau-z)(q\tau-z')^2},
$$
$$
J_{02}=\int\limits_{-1}^{1}\dfrac{\tau d\tau}{q\tau-z'}=
\dfrac{2}{q}+\dfrac{z'}{q^2}\ln\dfrac{z'-q}{z'+q}.
$$

Найдем числовую плотность (концентрацию) частиц плазмы,
отвечающую вырожденному распределению Ферми
$$
N=\int \Theta(1-P^2)\dfrac{2d^3p}{(2\pi\hbar)^3}=
\dfrac{2 p_0^3}{(2\pi\hbar)^3}\int \Theta(1-P^2)d^3P=
\dfrac{k_0^3}{3\pi^2},
$$
где $k_0$ -- волновое число Ферми, $k_0=\dfrac{mv_0}{\hbar}$.

В выражении перед интегралом из (2.4) выделим плазменную
(ленгмюровскую) частоту
$$
\omega_p=\sqrt{\dfrac{4\pi e^2N}{m}}
$$
и числовую плотность $N$,
и последнюю выразим через волновое число Ферми. Получим
$$
{j_x}^{\rm long}=\Big(\dfrac{e\Omega_p^2}{k_0p_0}\Big)
\dfrac{3k{E_y^2}}{32\pi\Omega q^3}\Big[z'J_{12}+\dfrac{2\gamma}{q} J_{02}\Big],
\eqno{(2.5)}
$$
где
$$
\Omega_p=\dfrac{\omega_p}{k_0v_0}=\dfrac{\hbar\omega_p}{mv_0^2}
$$
-- безразмерная плазменная частота.

Величину $ \gamma $ можно упростить, сводя к одномерным интегралам входящие
в $ \gamma $ интегралы (см. (1.15)). Заметим, что
$$
4\pi+\dfrac{iy}{2}J_0=4\pi+\pi iy
\int\limits_{-1}^{1}\dfrac{d\tau}{q\tau-z'}=
2\pi J_{01},
$$
где
$$
J_{01}=\int\limits_{-1}^{1}\dfrac{q\tau-\Omega}{q\tau-z'}d\tau=
2+\dfrac{iy}{2q}\ln\dfrac{z'-q}{z'+q}.
$$
Кроме того,
$$
J_1=\int\Bigg[qP_y\dfrac{\partial}{\partial P_x}
\Big(\dfrac{\partial f_0/\partial P_y}{qP_x-z}\Big)+
\dfrac{\Omega-qP_x}{qP_x-z}\cdot\dfrac{\partial^2f_0}{\partial P_y^2}\Bigg]
\dfrac{d^3P}{qP_x-z'}=
$$
$$
=q\int P_y\dfrac{\partial}{\partial P_x}
\Big(\dfrac{\partial f_0/\partial P_y}{qP_x-z}\Big)\dfrac{d^3P}{qP_x-z'}=
-q^2\int \dfrac{P_y[\partial f_0/\partial P_y]d^3P}{(qP_x-z)(qP_x-z')^2}=
$$
$$
=q^2\int \dfrac{f_0(P)d^3P}{(qP_x-z)(qP_x-z')^2}=
\pi q^2 \int\limits_{-1}^{1}
\dfrac{(1-\tau^2)d\tau}{(q\tau-z)(q\tau-z')^2}=
$$
$$
=\pi q^2J_{12}.
$$

Окончательно находим
$$
\gamma=\dfrac{(iy/2)J_1}{4\pi+(iy/2)J_0}=
\dfrac{iy}{4}q^2 \dfrac{J_{12}}{J_{01}},
$$
$$
z'J_{12}+\dfrac{2\gamma}{q}J_{02}=
\Big[z'+\dfrac{iy}{2}q\dfrac{J_{02}}{J_{01}}\Big]J_{12}
$$

Equality (2.5) we rewrite in the form
$$
j_x^{\rm long}=J(\Omega,y,q)\sigma_{l,tr}kE_y^2,
\eqno{(2.6)}
$$
где $\sigma_{l,tr}$ -- продольно--поперечная проводимость,
$J(\Omega,y,q)$ -- безразмерная плотность тока,
$$
\sigma_{l,tr}=
\dfrac{e \Omega_p^2}{p_0k_0}=\dfrac{e\hbar}{p_0^2}
\Big(\dfrac{\hbar \omega_p}{mv_0^2}\Big)^2=
\dfrac{e}{k_0p_0}\Big(\dfrac{\omega_p}{k_0v_0}\Big)^2,
$$
$$
J(\Omega,y,q)=\dfrac{3}{32\pi\Omega}
\Big[\Omega+\dfrac{iy}{2}\Big(1+q\dfrac{J_{02}}{J_{01}}\Big)\Big]J_{12}.
$$

Здесь
$$
q\dfrac{J_{02}}{J_{01}}=
\dfrac{2q+\Big(\Omega+\dfrac{iy}{2}\Big)\ln\dfrac{z'-q}{z'+q}}
{2q+\dfrac{iy}{2}\ln\dfrac{z'-q}{z'+q}}=
1+\dfrac{\Omega\ln\dfrac{z'-q}{z'+q}}{2q+
\dfrac{iy}{2}\ln\dfrac{z'-q}{z'+q}}.
$$

Если ввести поперечное поле
$$
\mathbf{E}_{\rm tr}=\mathbf{E}-\dfrac{\mathbf{k(Ek)}}{k^2}=
\mathbf{E}-\dfrac{\mathbf{q(Eq)}}{q^2},\qquad
{\bf kE}_{tr}=\dfrac{\omega}{c}[{\bf E,H}],
$$
то равенство (2.6) можно переписать в инвариантной форме
$$
\mathbf{j}^{\rm long}=J(\Omega,y,q)\sigma_{l,tr}{\bf k}{\bf E}_{tr}^2
=J(\Omega,y,q)\sigma_{l,tr}\dfrac{\omega}{c}[{\bf E,H}].
$$

{\sc \bf Замечание 1.}
Из формулы (2.5) (или (2.6)) видно, что при $y=0$ (или
$ \nu=0$) т.e. когда столновительная плазма переходит в
бесстолкновительную ($z\to \Omega, z '\to \Omega $),
эта формула в точности переходит в соответствующую формулу из
нашей работы \cite{Lat7} для бесстолкновитешльной плазмы
$$
{j_x}^{\rm long}=\sigma_{\rm l,tr}k{E_y^2}
\dfrac{3}{32\pi}\int\limits_{-1}^{1}
\dfrac{(1-\tau^2)d\tau}{(q\tau-\Omega)^3}.
$$

Перейдем к рассмотрению случая малых значений волнового числа.
Из выражения (2.6) при малых значениях волнового числа получаем
$$
{j_x}^{\rm long}=-\sigma_{\rm l,tr}k{E_y^2}
\dfrac{1}{8\pi \Omega (\Omega+iy)}.
$$

{\sc \bf Замечание 2.} При $y=0\; (\nu=0)$ из этой формулы в
точности вытекает соответствующая формула (1.15) из \cite{Lat7}
для продольного тока в случае малых значений волнового числа в
бесстолкновительной плазме.

\section*{\bf 3 \quad Заключение}

На рис. 1 и 2 представим поведение действительной (рис. 1)
и мнимой (рис. 2) частей плотности безразмерного продольного тока
при $ \Omega=1$ в зависимости от безразмерного волнового числа
$q $ при различных значениях безразмерной частоты столкновений электронов
с частицами плазмы.
При малых и больших значениях параметра $q$ кривые {\it 1,2} и {\it 3}
сближаются и становятся неразличимыми. Действительная часть
имеет сначала минимум, а затем максимум. С ростом безразмерной
частоты столкновений электронов мнимая часть плотности тока имеет один
максимум.

На рис. 3 и 4 представим поведение действительной (рис. 3) и мнимой
(рис. 4) частей плотности продольного тока в зависимости от
безразмерных волновых чисел  $q $ в случае $y=0.05$
при различных значениях безразмерной частоты колебаний
электромагнитного поля. Действительная часть сначала имеет
минимум, а затем максимум.
При больших значениях безразмерного волнового числа
кривые {\it 1,2} и {\it 3} сближаются и становятся
неразличимыми.

На рис. 5 и 6 представим поведение действительной (рис. 5) и
мнимой (рис. 6) частей плотности продольного тока в зависимости
от безразмерного волнового числа в случае $y=0.05$.
Действительная часть сначала имеет
минимум, а затем максимум. Мнимая часть имеет единственный
максимум.
При возрастании безразмерного волнового числа
$q $ кривые  {\it 1,2} и {\it 3} сближаются и практически совпадают.

В настоящей работе рассмотрено влияние нелинейного характера
взаимодействия электромагнитного поля с классической вырожденной
столкновительной плазмой.

Оказалось, что наличие нелинейности электромагнитного поля в
уравнении Власова
приводит к генерированию электрического тока, ортогонального к
направлению поля.

В дальнейшем авторы намерены рассмотреть задачу плазменных
колебаний и задачу о скин-эффекте с использованием квадрата
векторного потенциала в разложении функции распределения.

\clearpage

\begin{figure}[t]\center
\includegraphics[width=16.0cm, height=9cm]{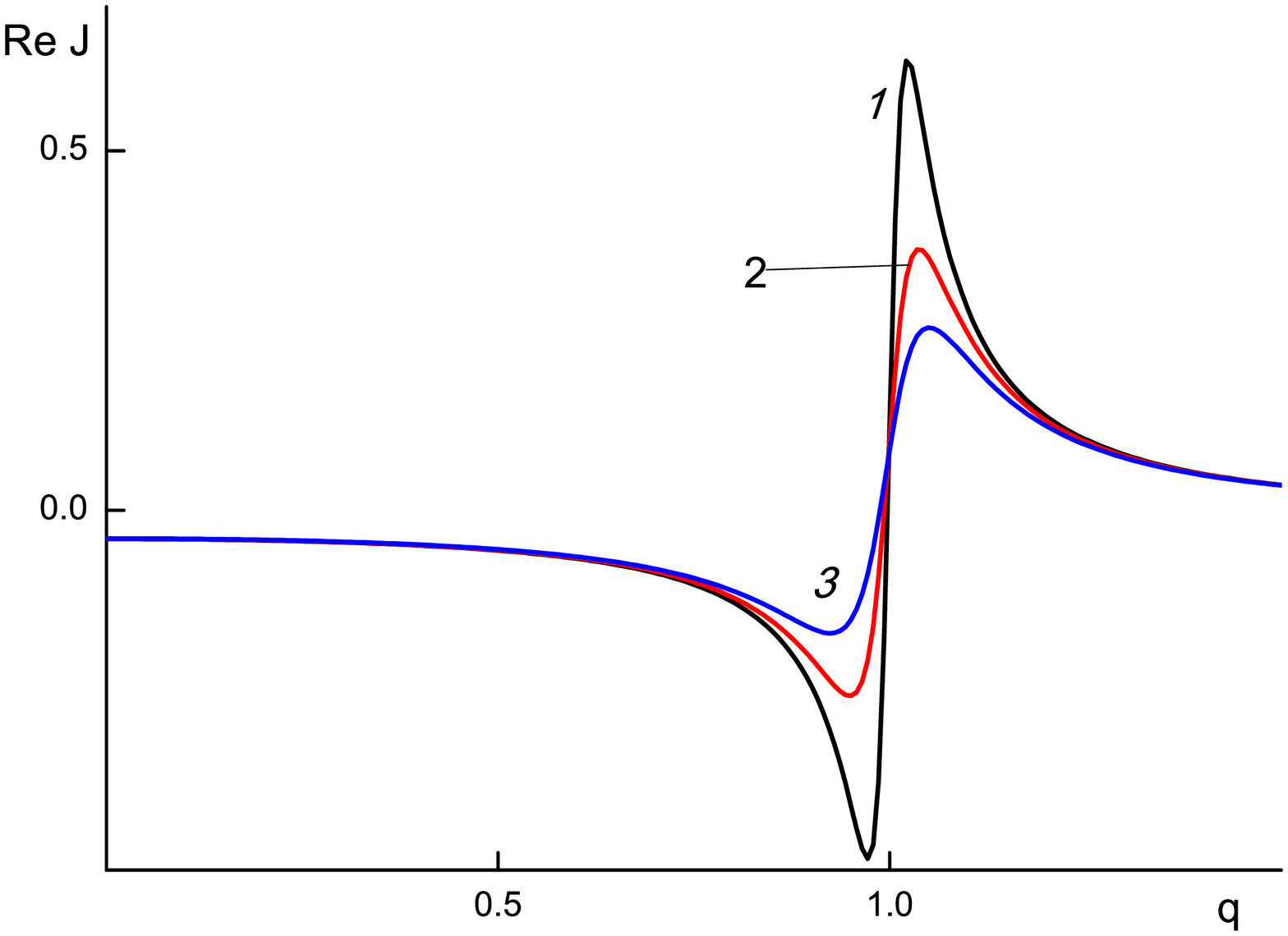}
{Рис. 1. Действительная часть плотности безразмерного продольного тока,
$\Omega=1$. Кривые $1,2,3$ отвечают значениям
безразмерной \\частоты столкновений $y=0.04, 0.07, 0.1$.}
\includegraphics[width=17.0cm, height=9cm]{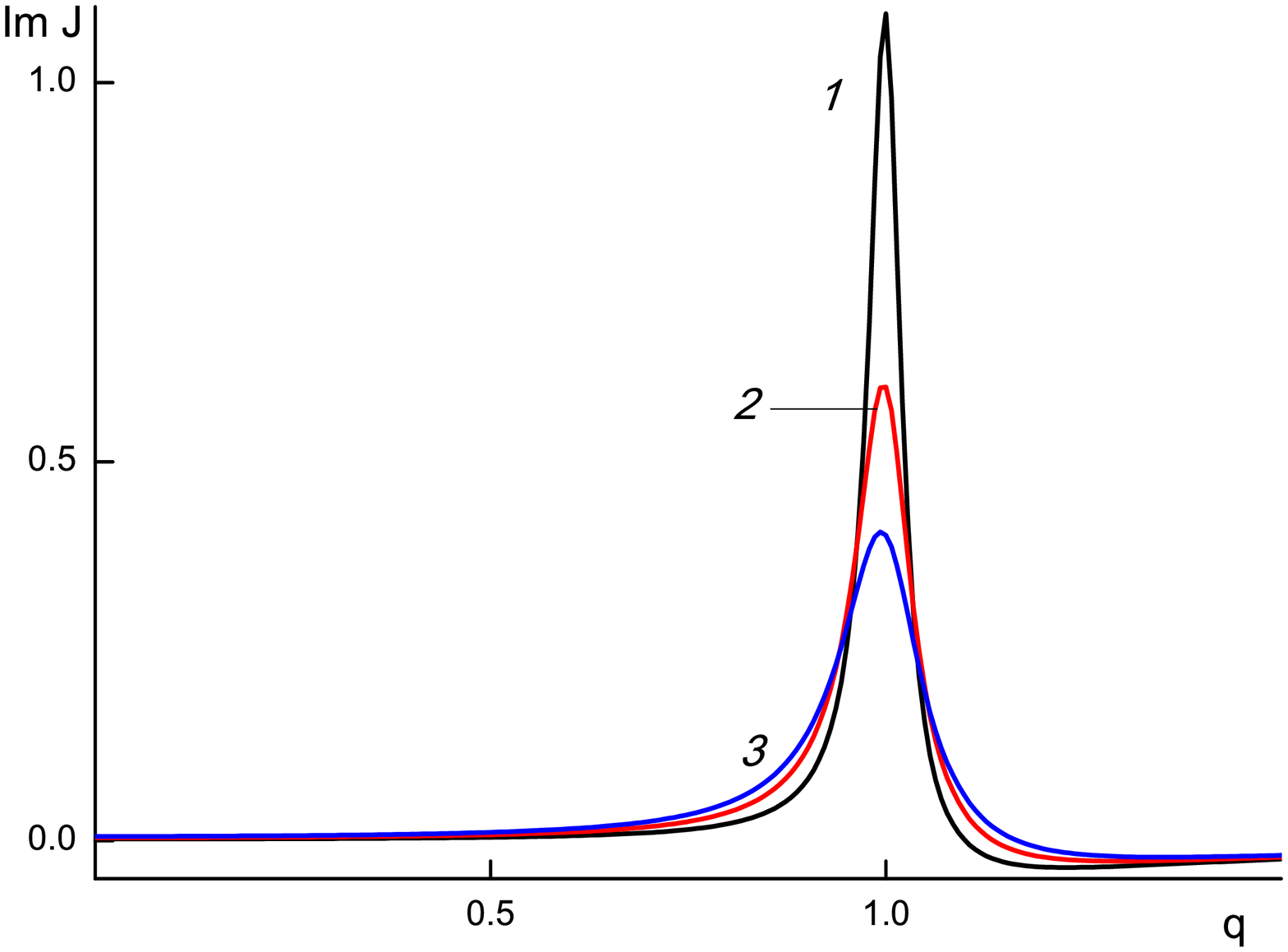}
\center{Рис. 2. Мнимая часть плотности безразмерного продольного тока,
$\Omega=1$. Кривые $1,2,3$ отвечают значениям
безразмерной частоты столкновений $y=0.04, 0.07, 0.1$.}
\end{figure}

\begin{figure}[ht]\center
\includegraphics[width=16.0cm, height=9cm]{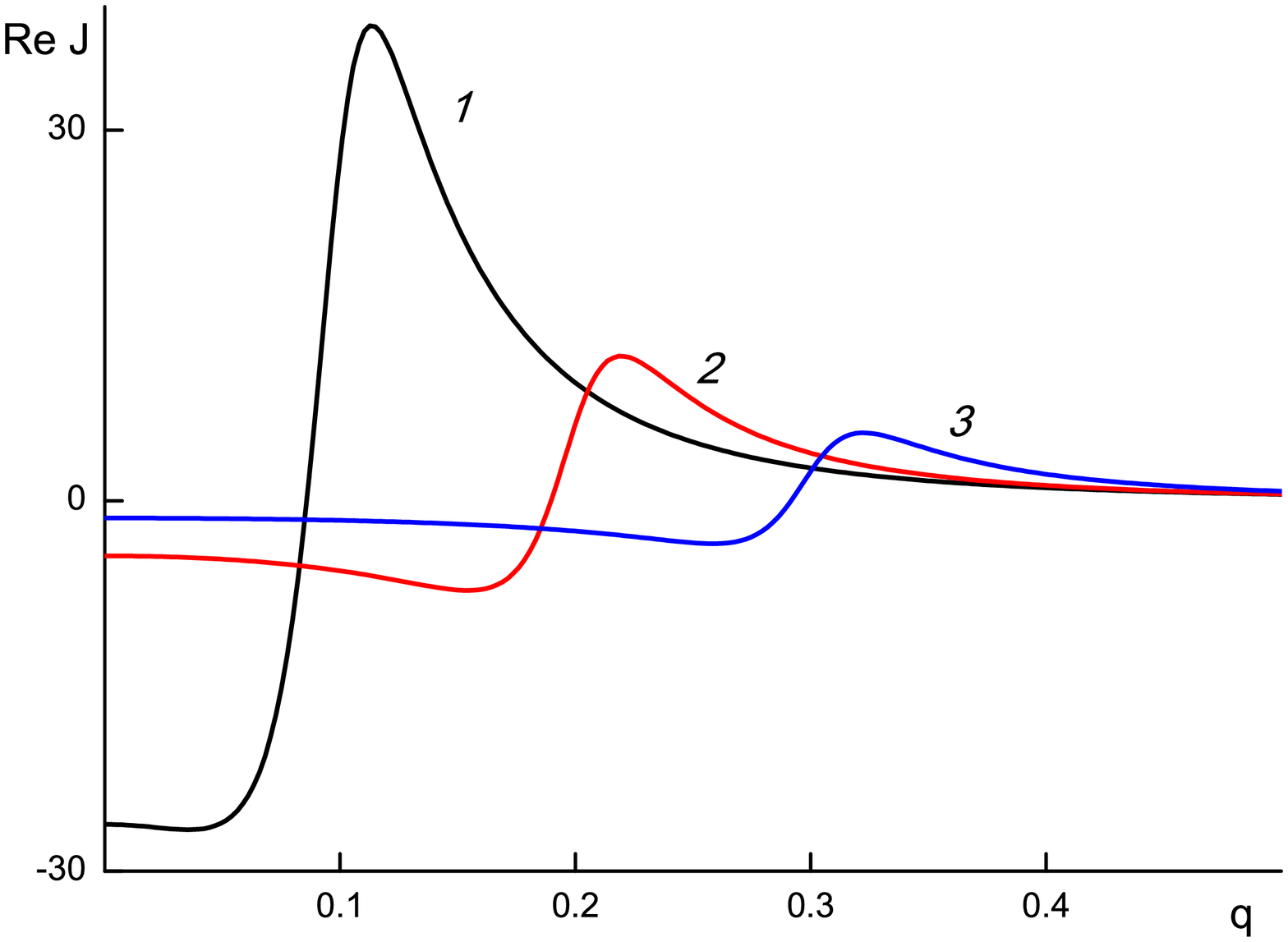}
\center{Рис. 3. Действительная часть плотности безразмерного продольного тока,
$y=0.05$. Кривые $1,2,3$ отвечают значениям
безразмерной частоты \\колебаний электромагнитного поля
$\Omega=0.1, 0.2, 0.3$.}
\includegraphics[width=17.0cm, height=9cm]{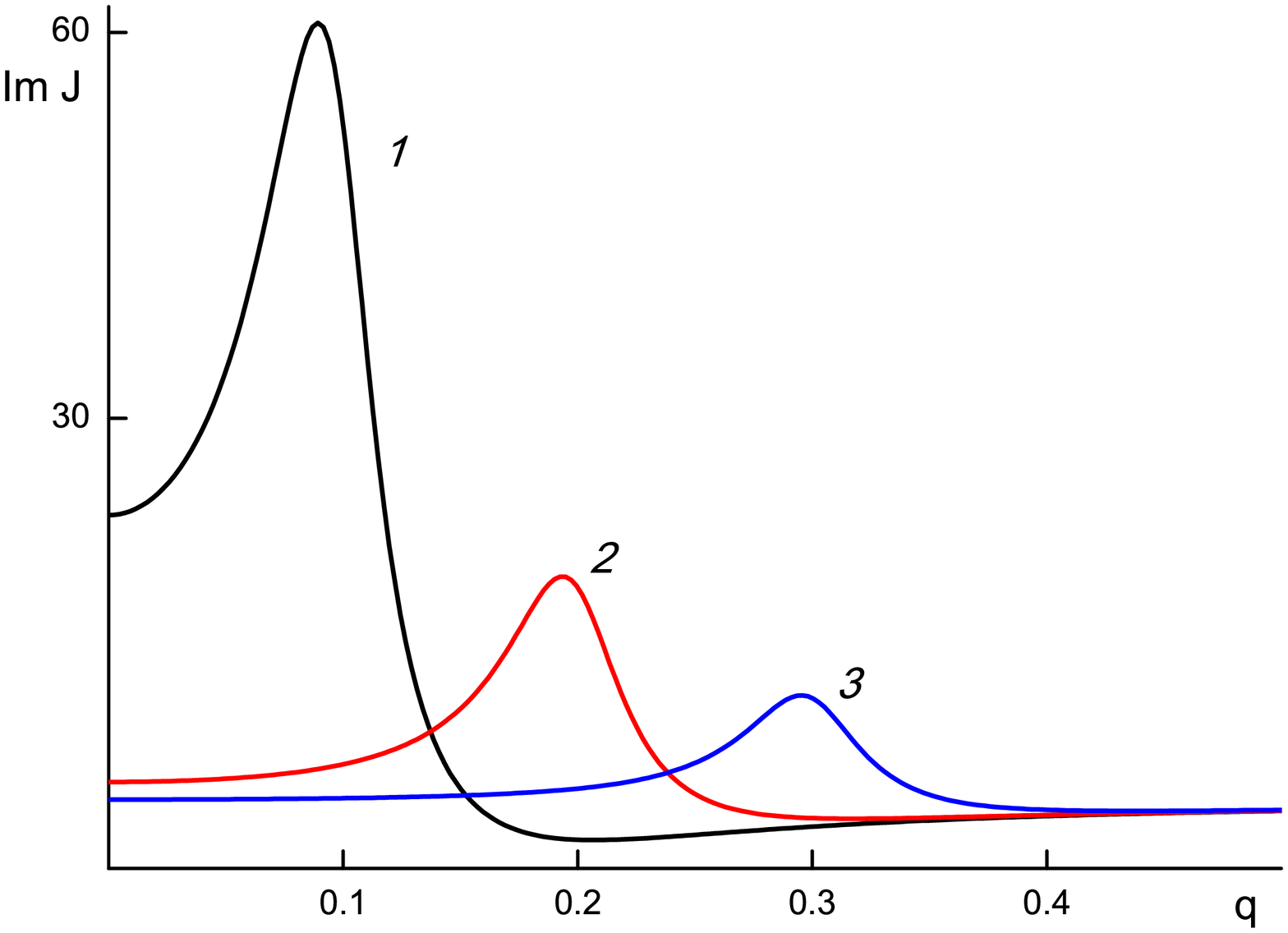}
\center{Рис. 4. Мнимая часть плотности безразмерного продольного тока,
$y=0.05$. Кривые $1,2,3$ отвечают значениям безразмерной частоты
колебаний электромагнитного поля  $\Omega=0.1, 0.2, 0.3$.}
\end{figure}

\begin{figure}[ht]\center
\includegraphics[width=16.0cm, height=9cm]{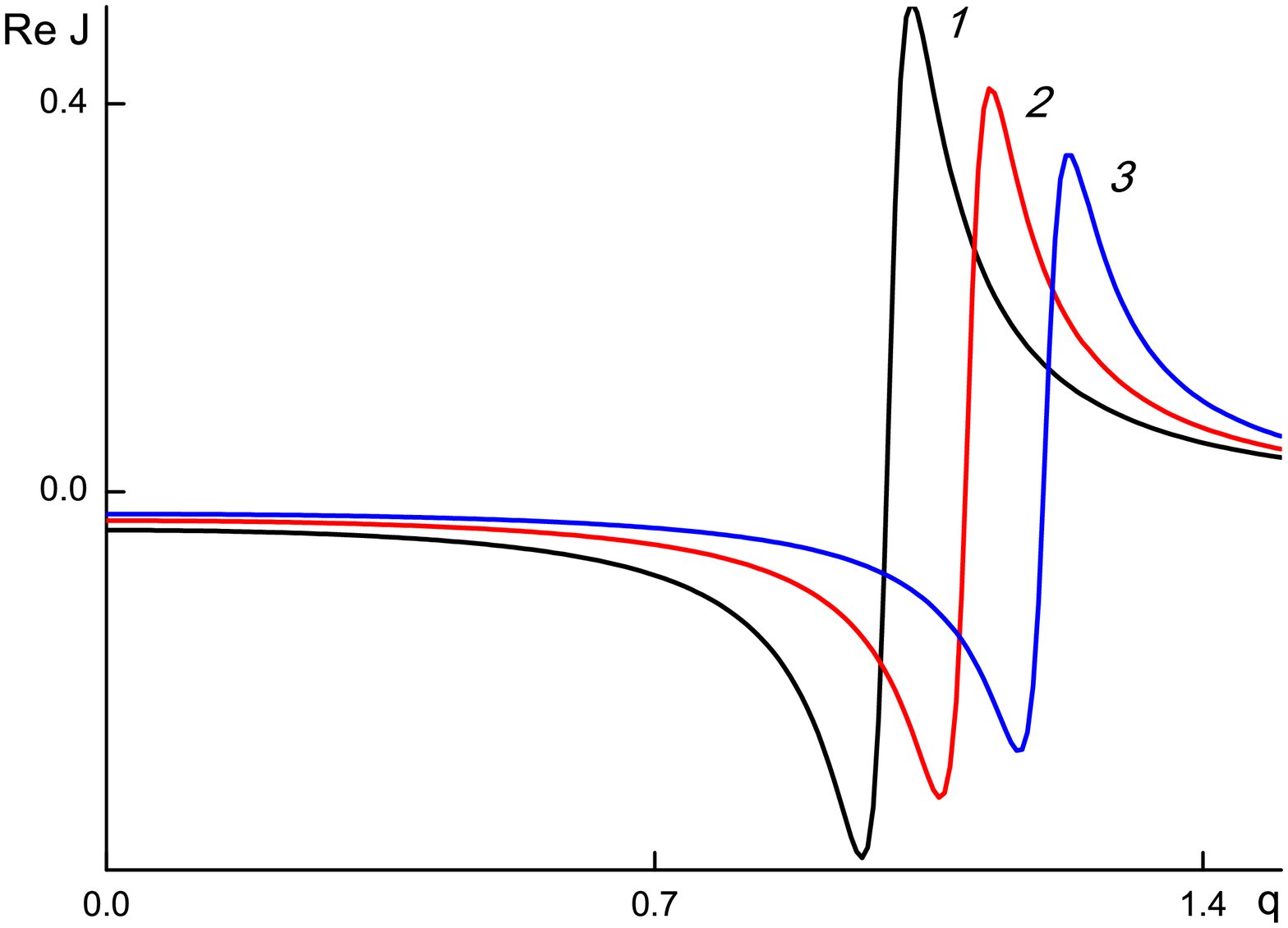}
{Рис. 5. Действительная часть плотности безразмерного продольного тока,
$y=0.05$. Кривые $1,2,3$ отвечают значениям
безразмерной частоты колебаний электромагнитного поля $\Omega=1, 1.1, 1.2$.}
\includegraphics[width=17.0cm, height=9cm]{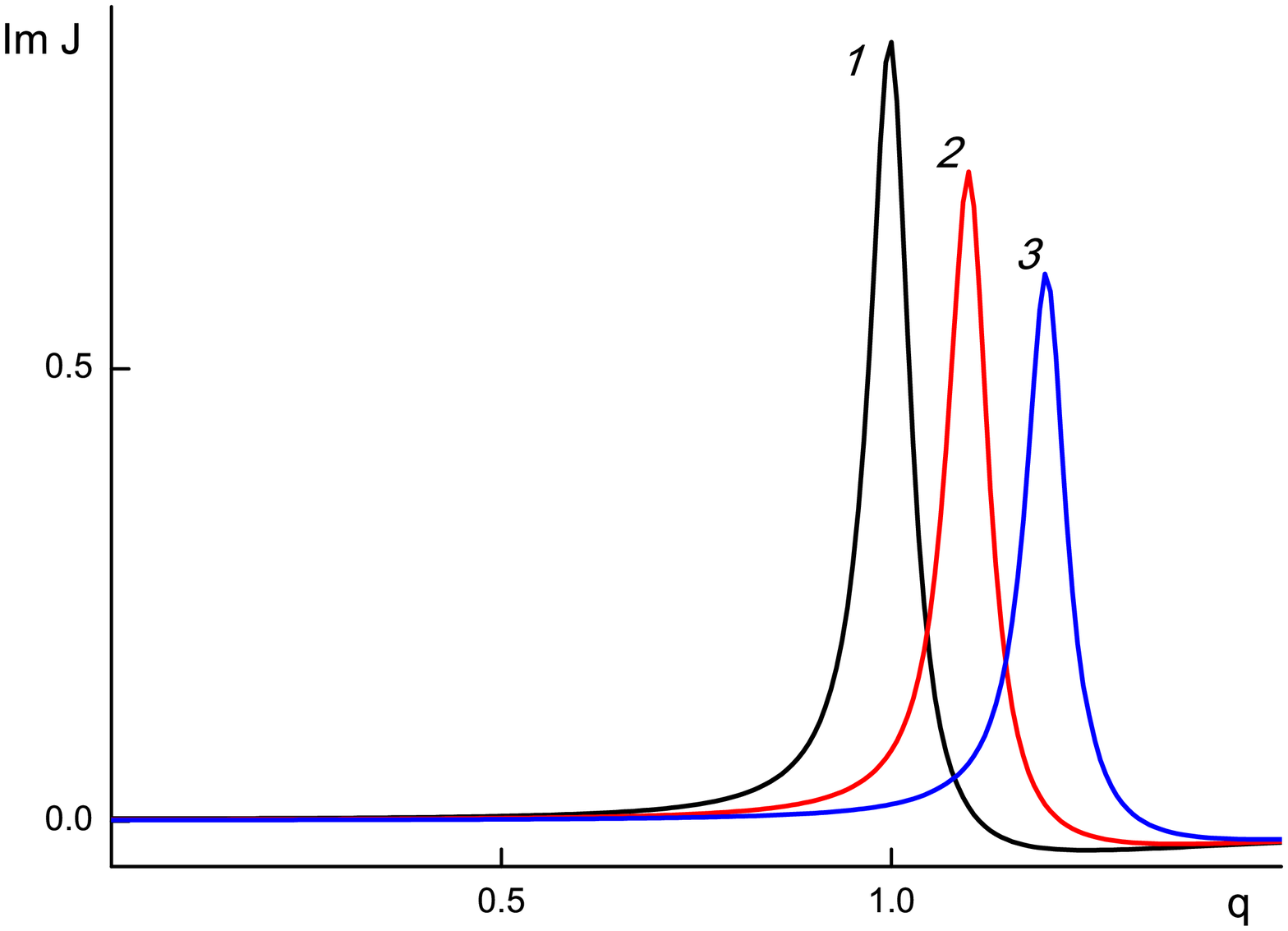}
\center{Рис. 6. Мнимая часть плотности безразмерного продольного тока,
$y=0.05$. Кривые $1,2,3$ отвечают значениям
безразмерной частоты колебаний электромагнитного поля $\Omega=1, 1.1, 1.2$.}
\end{figure}

\end{document}